\RequirePackage[2020-02-02]{latexrelease}
\documentclass[12pt,aps,superscriptaddress]{revtex4}

\usepackage{multirow}
\usepackage{amsmath,amsfonts}
\usepackage{amssymb,latexsym}
\usepackage{color}
\usepackage{graphicx}
\usepackage{mathrsfs}
\usepackage[normalem]{ulem}

\newtheorem{prop}{Proposition}

\newcommand{\beprop}{\begin{prop}}
\newcommand{\enprop}{\end{prop}}
\newcommand{\bprf}{\begin{proof}} 
\newcommand{\eprf}{\end{proof}\qed}

\definecolor{hervecolor}{rgb}{0.8,0,0.7}

\newcommand{\ket}[1]{|\kern.3ex#1\kern.3ex\rangle}
\newcommand{\bra}[1]{\langle\kern.3ex #1 \kern.3ex|}
\newcommand{\scalar}[2]{\langle\kern.3ex #1 \kern.3ex|\kern.3ex#2\kern.3ex\rangle}

\newcommand{\ii}{\mathsf{i}}

\newcommand{\ud}{\mathrm{d}}

\newcommand{\uE}{\mathrm{E}}

\begin{document}

\title{Unification of Stochastic and Quantum Thermodynamics in Scalar Field Theory via a Model with Brownian Thermostat
}
\author{T.\ Koide}
\email{tomoikoide@gmail.com,koide@if.ufrj.br}
\affiliation{Instituto de F\'{\i}sica, Universidade Federal do Rio de Janeiro, 
21941-972, Rio de Janeiro, RJ, Brazil}
\affiliation{Frankfurt Institute for Advanced Studies (FIAS), Frankfurt am Main, Germany}  
\author{F.\ Nicacio}
\affiliation{Instituto de F\'{\i}sica, Universidade Federal do Rio de Janeiro, 
21941-972, Rio de Janeiro, RJ, Brazil}

\begin{abstract}
We present a systematic procedure to derive a quantum master equation for thermal
relaxation in real scalar field theory, expanding on the method
proposed in [Koide and Nicacio, Phys. Lett. A494, 129277 (2024)]. 
We begin by introducing a generalized model for a 
classical scalar field interacting with a Brownian thermostat, consistent with stochastic thermodynamics.
Applying canonical quantization to this model, we derive the corresponding quantum
master equation, that is applicable to any form of the scalar field Hamiltonian.
While its evolution is generally non-CPTP (Completely Positive and Trace-Preserving), 
it can be adjusted to describe a 
CPTP evolution, such as those found in the GKSL (Gorini-Kossakowski-Sudarshan-Lindblad)
equation by appropriately tuning the parameters of the model.
In this framework, we define heat, work, and entropy in a way that satisfies the first 
and second laws of quantum thermodynamics. This suggests that the quantum-classical 
correspondence extends beyond closed systems governed by unitary time evolution to open 
systems as well. We further investigate the relation between the second law in 
quantum thermodynamics and relative entropy, providing insights into the study of quantum 
fluctuations through information-theoretical techniques in quantum field theory.
\end{abstract}
\maketitle

\section{Introduction}

Understanding thermodynamics in small fluctuating systems remains a central challenge in 
modern physics. Two primary frameworks address this challenge: stochastic thermodynamics, 
which applies to classical mesoscopic systems, and quantum thermodynamics, 
which is crucial 
for systems dominated by quantum fluctuations. 
The former expands standard thermodynamics by redefining heat, work, and entropy, while preserving 
the first and second laws, often using models based on Brownian motion
\cite{sekimoto,seifert,peliti}. However, classical approaches fail to describe quantum 
systems where fluctuations arise intrinsically from the principles of quantum mechanics. 
Quantum thermodynamics builds on the foundations of quantum physics \cite{qt1,qt2}, 
but the relation between its classical limit and stochastic thermodynamics has not
yet been well investigated. Our recent studies \cite{KoiNic2023,KoiNic2024} have hinted 
at a quantum-classical correspondence that bridges these two frameworks.

In quantum thermodynamics, thermal relaxation processes are typically modeled using the 
Gorini-Kossakowski-Sudarshan-Lindblad (GKSL) equation, which describes Completely Positive and 
Trace-Preserving (CPTP) dynamics. 
A significant challenge lies in determining the Lindblad's jump operators directly from
first principles. While quantum master equations can be derived through coarse-graining 
(the choice of gross variables and the Markov approximation) from 
microscopic dynamics, ensuring CPTP evolution requires additional approximations, such as 
the rotating wave approximation and the factorization in the initial system-bath density matrix 
\cite{BreuerPetr2002}.
There are numerous studies that attempt to advance research in this direction \cite{abbruzzo,becker,davidovic,pradilla,breuer04,gneiting,piilo,whitney,fogedby}. 
However, in this paper,  we explore an alternative approach.


The present authors recently proposed a novel approach to constructing quantum master equations, 
deviating from conventional methods commonly used in many-body systems \cite{KoiNic2023}. This 
approach aims to elucidate the role of quantum fluctuations in thermal relaxation processes by 
formulating a quantum master equation with a well-defined classical limit. Notably, in the 
Caldeira-Leggett model, the quantum counterpart of standard Brownian motion fails to yield a GKSL 
equation \cite{BreuerPetr2002}.
To address this issue, we first developed a generalized model of Brownian motion, 
where the relation between a particle's velocity and momentum is modified due to interactions with heat baths. 
We then demonstrated that this model provides a consistent framework for stochastic thermodynamics. By applying canonical quantization to the generalized model, we derived a quantum master equation. 
Our derivation is phenomenological; consequently, the coefficients in our equation are set by external inputs. Just as the viscous coefficients in hydrodynamics are determined through such means, these parameters should ideally be derived from microscopic theories. For instance, the projection operator technique might be employed to obtain the Green–Kubo–Nakano formula for these coefficients, as illustrated in Refs.\ \cite{zwanzig,koidekodama}.
It is worth noting that the canonical quantization of standard Brownian motion has been 
investigated by Oliveira \cite{oliveira1,oliveira2}. 
However, the generalization of Brownian motion is not considered and thus resulting equations do not satisfy the CPTP requirement.

The approach described in Ref.\ \cite{KoiNic2023} offers two key benefits that are fundamental to the extension of thermodynamics.
First, it provides simplicity: the quantum master equation is obtained solely by the system Hamiltonian and coupling coefficients, without requiring detailed knowledge of the system-bath interactions. In particular, Lindblad's jump operators and the modifications to the system Hamiltonian are automatically specified by 
these quantities.
Second, it ensures that the classical limit of the quantum master equation consistently describes thermal relaxation processes. This includes not only convergence to equilibrium but also adherence to 
inequalities involving heat and entropy in accordance with the second law of thermodynamics.
Although dynamics described by our quantum master equation are generally non-CPTP, they can be tuned to match the GKSL equation for general quadratic  
many-body Hamiltonians by appropriately adjusting the coupling parameters \cite{KoiNic2023,KoiNic2024}.

In the present paper, we will put forward an extension of our program \cite{KoiNic2023,KoiNic2024} to a field theoretical model. 
This may be useful in furthering our understanding of the interplay between 
quantum field and quantum information theories.
Quantum field theory employs various entropy measures, such as entanglement entropy, relative entropy, and mutual
information to characterize the intricate nature of quantum fluctuations \cite{nishioka, witten}. 
To bridge these information-theoretic entropies with thermodynamic entropy, 
it is essential to establish a framework for quantum thermodynamics within field theory. 
Moreover, investigating the relation between thermalization and time-reversal symmetry breaking becomes particularly intriguing in field-theoretical models of open systems, 
such as K mesons, where CP symmetry is violated. 
Recent proposals also suggest that GKSL equations may describe the interaction of
quantized matter fields with classical gravity, further underscoring the relevance of this
framework \cite{oppen}.
Furthermore, several studies in cosmology have employed field-theoretical GKSL equations \cite{burgess, Boya, hollow, martin, brahma, colas,alicki23}. 
To establish a systematic procedure for deriving such a GKSL equation, it is important to examine its derivation 
from various perspective.

With few exceptions, research in stochastic thermodynamics has focused exclusively on cases where the number of particles remains constant, 
thus neglecting the contribution from chemical potential.
One such attempt is the formulation in Ref.\ \cite{sekimoto2000}, which is a particle-based standard approach that uses the positions and momenta of particles as its dynamic variables. 
While this allows for a detailed description of the system in classical phase space, attempting to model changes in particle number from this microscopic level results 
in a highly complex formulation.
In contrast, the framework in Ref.\ \cite{schmiedl} is a mesoscopic model designed specifically to handle chemical reactions interacting with a chemiostat. 
To achieve this, the model defines the system's state by the set of the numbers of particles, 
thereby forgoing a description in terms of classical phase-space variables by assuming spatial homogeneity and negligible interaction energy. 
Differently from Ref.\ \cite{sekimoto2000}, the stochasticity is characterized by a Markovian jump process of the set of the numbers of particles.
A formulation based on field theory, as we propose, may provide a path to unify these perspectives, 
offering a framework capable of describing systems with both detailed spatial configurations and variable particle numbers. 
This provides another motivation for discussing the extension to field theory.

In this paper, we systematically derive a quantum master equation that describes thermal 
relaxation processes in real scalar field theory.
We begin by introducing a classical field-theoretical model where a scalar field interacts with a Brownian thermostat, ensuring consistency with stochastic thermodynamics. By applying canonical quantization, we derive a quantum master equation applicable to arbitrary scalar-field Hamiltonians.
Although this equation generally describes non-CPTP evolution, we show that, for a free scalar field Hamiltonian, the parameters can be adjusted to ensure CPTP evolution. Within this framework, heat, work, and entropy are defined in a manner consistent with the first and second laws of quantum thermodynamics. 
Furthermore, our approach guarantees that quantum thermodynamics converges to stochastic thermodynamics in the classical limit ($\hbar \to 0$).
Finally, we examine the distinctions between our formulation of the second law 
and the concept of relative entropy.

This paper is organized as follows.
In Sec. \ref{sec:math}, we discuss the discretization of scalar fields to further introduce stochastic behaviors in these fields.
Using this result, we introduce the classical field-theoretical model interacting with Brownian thermostat and demonstrate that the model satisfies the first and second law in stochastic thermodynamics in Sec.\ \ref{sec:stother}. 
Our quantum master equation is obtained by applying canonical quantization to our classical model.
Choosing the free field as the system Hamiltonian, 
we show that the derived dynamics is a CPTP map. 
Heat, work, and entropy are introduced by generalizing the corresponding quantity in stochastic thermodynamics showing the first and second laws in quantum thermodynamics in Sec.\ \ref{sec:quather}.
In Sec.\ \ref{sec:rel_ent}, 
the relation between the second law of quantum thermodynamics and the entropy production defined through the relative entropy is discussed. 
Section \ref{sec:concl} is devoted to concluding remarks.

\section{Mathematical set up and Brownian Thermostat} \label{sec:math}

As well-known, the trajectory of a Brownian particle is not differentiable by the influence of noise in its temporal evolution \cite{book:gardiner}.
To describe a fluctuating field configuration, the field at each point in spacetime is subject to random fluctuations due to noise terms. 
Therefore, a field configuration at each stochastic event is a non-differentiable function. To handle such a random effect, 
it is necessary to introduce the discretization of field in spacetime.
The discretization method of fields is an extension of the work developed by one of the present authors \cite{koide_ptp}.

Let us consider the $1+1$-dimensional system of total length $\ell$, where 
positions are represented by $2N$ discretized points 
in a grid with size $\Delta x = \ell/(2N)$, that is,
$x_i = i \Delta x$ for $i=-N,-(N-1),\cdots, N-1$.
Assuming, for convenience, the periodic boundary conditions for a bosonic field $\psi(x_i)$, we have
\begin{align}
\psi(x_{i + 2N}) &= \psi(x_i) \, .
\end{align}
Therefore the minimum is located at $x_{\rm min} = x_{-N} = -\ell/2$ and the maximum is at $x_{\rm max} = x_{N-1} = \ell/2 - \Delta x$.  
Due to the periodic boundary condition, the field at $x_N = \ell/2$ is defined through 
$\psi(x_{N}) = \psi(x_{-N})$ and thus the field at $x_{N}$, $\psi(x_{N})$, 
is not explicitly considered in the following calculation.

Because of this discretization, the spatial derivatives are replaced by differences:
the first and second derivatives are, respectively, given by  
\begin{align}
\partial_x &= \frac{1}{2\Delta x}
\begin{pmatrix}
		0 & 1 & 0 & \cdots & 0 & -1 \\
		-1 & 0 & 1 & \cdots & 0 & 0\\
		0 & -1 & 0 & \cdots & 0 & 0 \\
		\vdots & & & & & \vdots \\
		 0 & \cdots & \cdots & -1 & 0 & 1\\
		 1 & 0 & \cdots & 0 & -1 & 0
	\end{pmatrix} \, , \,\,\,\, 
\partial^2_x &= \frac{1}{(2\Delta x)^2}
\begin{pmatrix}
		-2 & 0 & 1 & \cdots & 1 & 0 \\
		0 & -2 & 0 & \cdots & 0 & 1\\
		1 & 0 & -2 & \cdots & 0 & 0 \\
		\vdots & & & & & \vdots \\
		 1 & \cdots & \cdots & 0 & -2 & 0\\
		 0 & 1 & \cdots & 1 & 0 & -2
\end{pmatrix}\, .
\end{align}
These definitions are different from those in Ref.\ \cite{koide_ptp}.

Let us denote the eigenvectors of $\partial^2_x$ as ${u}^{(k_n)}$ associated to the eigenvalues $\lambda^2_{k_n}$,
\begin{align}
\partial^2_x {u}^{(k_n)} = - \lambda^2_{k_n} {u}^{(k_n)} \, , 
\end{align}
where
\begin{align}
\lambda_{k_n} = \frac{\sin (k_n \Delta x)}{\Delta x} \label{eqn:lambda}\, , 
\end{align}
and $k_n = 2\pi n/\ell$ for $n=-N, -(N-1), \cdots, N-1$.
These eigenvectors are orthonormal and constitute a complete-set, i.e.,
\begin{align}
\sum_{i=-N}^{N-1} ({u}^{(k_n)} (x_{i}))^\dagger {u}^{(k_m)} (x_{i}) 
&= \frac{1}{\Delta x} \delta_{n,m} 
\label{eqn:orthocond}
\, ,\\
\Delta x \sum_{n=-N}^{N-1}({u}^{(k_n)} (x_\alpha))^\dagger {u}^{(k_n)} (x_\beta) 
&= \delta_{\alpha,\beta}\, ,
\label{eqn:completeset}
\end{align}
where
\begin{align}\label{eqn:eigvec}
{u}^{(k_n)} (x_{i})
&= \frac{1}{\sqrt{\ell}}e^{-\ii k_n x_i} \, ,\\
({u}^{(k_n)})^T 
&=
\left(
{u}^{(k_n)}(x_{-N}), {u}^{(k_n)}(x_{-N+1}), \cdots, {u}^{(k_n)}(x_{N-1})
\right) \label{eq:base}\, .
\end{align}

In the following, we consider a system described by a real scalar field. 
The real scalar field and its conjugate field can be expanded as
\begin{align}
\Phi_t (x_i) &= \sqrt{\Delta k} \sum_{n=-N}^{N-1} \phi _t (k_n) u^{(k_n)} (x_i) \, , 
\label{eqn:Phi_phi}\\
\pi_t (x_i) &= \sqrt{\Delta k} \sum_{n=-N}^{N-1} \Pi_t (k_n) u^{(k_n)} (x_i) \, , 
\label{eqn:pi_Pi}
\end{align}
where $\Delta k = 2\pi/\ell$. 
For simplicity in notation, we use $A_t(x) = A(x, t)$ to represent a general time-dependent function. 
From now on, we will develop the formulation in terms of 
$\phi (k_n)$ and $\Pi (k_n)$. 
For a given Lagrangian $L$ the conjugate field is given by 
\begin{equation}
\Pi_t (k_n) = \frac{1}{\Delta k} \frac{\partial L}{\partial \dot{\phi}_t (-k_n)} \, ,
\end{equation}
where the symbol ``$\,\dot{\bullet}\,$''
denotes the temporal derivative and 
$L = L({\Phi_t (x), \dot{\Phi}_t (x)})$ is the discretized Lagrangian.
To represent a family of fields, the notation 
$\{A_t(x), B_t(x), \cdots\}$ for arbitrary fields 
$A_t(x), B_t(x), \cdots$ refers to the discrete set 
$\{A_t(x_i), B_t(x_i) \mid i = -N, \cdots, N-1\}$.

From a Legendre transformation, we introduce the  Hamiltonian,
\begin{align}
H 
&= \Delta k \sum_n \left\{
\Pi_t(k_n) \dot{\phi}_t(-k_n) - L 
\right\}  \, ,
\end{align}
and then the corresponding canonical equations describe the behavior of an isolated system,  
\begin{align}
\partial_t \phi_t(k_n) &= \frac{1}{\Delta k}\frac{\partial H}{\partial \Pi_t(-k_n)} \, , \\
\partial_t \Pi_t(k_n) &=-  \frac{1}{\Delta k}\frac{\partial H}{\partial \phi_t(-k_n)}  \, .
\end{align}

We further summarize
the expansion of noise terms which appear in stochastic differential equations.
The symbol ``\,$\widehat{\bullet}$\,'' 
will be employed to denote a stochastic quantity. 
Let us introduce the Wiener process $ \widehat{B}_t (x_i)$ in the discretized spacetime 
satisfying
the following correlations:
\begin{align}
\uE \left[ \ud\widehat{B}_t (x_i) \right] 
&= 0 \, , \label{eq:mean_Bx}\\
\uE \left[ \ud\widehat{B}_t (x_i)  \ud\widehat{B}_{t^\prime} (x_j)\right] 
&= \ud t \, \delta_{t,t^\prime} \delta_{i,j}  \, , \label{eq:var_Bx}
\end{align}
where $\ud \widehat{B}_t(x_i) := \widehat{B}_{t+\ud t}(x_i) - \widehat{B}_t(x_i)$
is the increment and $\uE \left[ \bullet \right]$ is the expectation value.
As before, we will consider the expansion of $\ud\widehat{B}_t (x_i)$
using the complete set where the expansion coefficients are denoted by
$\ud\widehat{\cal B}_t (k_n)$, {\it i.e.},
\begin{equation}
\ud\widehat{B}_t(x_i) = \sqrt{\Delta k} \sum_{n=-N}^{N-1} \left( \sqrt{\frac{\Delta x}{\Delta k}} \ud\widehat{\cal B}_t (k_n) \right) u^{(k_n)} (x_i) \, .
\end{equation}
The correlation properties of $\ud\widehat{\cal B}_t (k_n)$ are obtained using the above expansion in Eqs.\ (\ref{eq:mean_Bx}) and (\ref{eq:var_Bx}) with Eqs.\ (\ref{eqn:orthocond}) and (\ref{eqn:completeset}), namely,
\begin{align}
\uE \left[ \ud\widehat{\cal B}_t (k_n) \right]
&= 0 \, , \label{eq:corr_dB1}\\ 
\uE \left[ \ud\widehat{\cal B}_t (k_n)  \ud\widehat{\cal B}_{t^\prime} (-k_m)\right]
&= \ud t \, \delta_{t,t^\prime} \delta_{n,m}  \, .  \label{eq:corr_dB2}
\end{align}

Now, we proceed to determine the stochastic differential equation for the field interacting with Brownian thermostat. 
Let us first consider an isolated Hamiltonian system, the form of which can be controlled by time-dependent external parameters.
Since we consider a field-theoretical system, such a parameter is a function of position in general.
As the simplest example, we choose a Hamiltonian depending on a single external field $b_t (k_n)$, which satisfies periodic boundary conditions.
We simply assume that the time dependence of $b_t (k_n)$ is given as an externally prescribed function of time.
Then the system Hamiltonian of the real scalar field is denoted by 
$H(\{{\Pi}_t,{\phi}_t, b_t \})$.
The Brownian thermostat has a fixed temperature $T = 1/ (k_{\rm B} \beta)$ with $k_{\rm B}$ 
being the Boltzmann constant.
Due to the interaction with the Brownian thermostat, 
the scalar field dynamics becomes dissipative and is influenced by thermal noise.
The fields $\Pi_t(k_n)$ and $\phi_t(k_n)$ are then not differentiable and are defined only on discretized momenta and times.

Extending our previous results in Ref.\ \cite{KoiNic2023} for many-body systems,
a field-theoretical model interacting with Brownian thermostat is supposed to be described by the following stochastic differential equations:
\begin{align}
\ud \widehat{\phi}_t(k_n) &=  \frac{\ud t}{\Delta k} \left\{ \frac{\partial }{\partial \widehat{\Pi}_t(-k_n)} 
- \gamma_\phi \frac{\partial}{\partial \widehat{\phi}_t(-k_n)}  \right\}
H(\{\widehat{\Pi}_t,\widehat{\phi}_t, b_t\})
+ \sqrt{\frac{2\gamma_\phi }{\Delta k \, \beta} } \ud \widehat{\cal B}^\phi_t(k_n)\, ,  \label{eqn:sde1} \\
\ud \widehat{\Pi}_t(k_n) &= - \frac{\ud t}{\Delta k} \left\{  \frac{\partial }{\partial \widehat{\phi}_t(-k_n)}  
+ 
\gamma_\Pi \frac{\partial}{\partial \widehat{\Pi}_t(-k_n)} 
\right\}H(\{\widehat{\Pi}_t,\widehat{\phi}_t,b_t\})
+ \sqrt{\frac{2\gamma_\Pi}{\Delta k \, \beta}} \ud \widehat{\cal B}^\Pi_t(k_n) \, , \label{eqn:sde2}
\end{align}
where $\ud \widehat{A}_t := \widehat{A}_{t+\ud t} - \widehat{A}_t$ and 
$\ud t$ denotes the width of the discretized time.
The coefficients $\gamma_\phi = \gamma_\phi(|k_n|,t)$ and 
$\gamma_\Pi = \gamma_\Pi(|k_n|,t)$ are arbitrary real and positive functions of $|k_n|$ and $t$. 
Again, we simply assume that  these time dependences are given as externally prescribed functions of time.
Note that we employed the fluctuation-dissipation theorem (Einstein relation) to express the noise intensities as function of $\gamma_\mu$. For further details, see Ref.\ \cite{KoiNic2023}.
In the above complex representations of the stochastic differential equations, 
we introduce $\ud \widehat{\cal B}_t^\mu(k_n)$ with $\mu = \phi, \Pi$, 
which satisfy the same correlations defined by Eqs.\ (\ref{eq:corr_dB1}) and  (\ref{eq:corr_dB2}), 
but are independent of each other,
\begin{align}
\begin{split} \label{eqn:correlations_dbk}
\uE\left[ \ud \widehat{\cal B}_t^\mu (k_n)  \right] & = 0 \, ,\\
\uE \left[ \ud \widehat{\cal B}_t^\mu (k_n) \, \ud \widehat{\cal B}^\nu_{t'} (-k_m) \right]
&= \ud t \, \delta_{\mu,\nu} \delta_{m,n} \delta_{t,t^\prime} \,.
\end{split}
\end{align}

The field configuration is characterized by the probability distribution defined by 
\begin{align}
\rho_t(\{\Pi, \phi\}) 
&=\int [\ud \Pi_0] \int [\ud \phi_0] \, \rho_0(\{\Pi_0, \phi_0 \}) 
\prod^{N-1}_{n=-N}\uE \left[\delta (\Pi (k_n) - \widehat{\Pi}_t(k_n)) 
\delta (\phi (k_n) - \widehat{\phi}_t(k_n)) \right] 
\, , \label{eqn:field_conf}
\end{align} 
where $[\ud A] := \prod_{n=-N}^{N-1} \ud A (k_n)$, 
$\phi_0$ and $\Pi_0$ are the corresponding fields at a given initial time $t_0$ and $\rho_0(\{\Pi_0, \phi_0 \})$ represents the initial probability distribution of these fields, normalized by one.

Assuming that this distribution converges in the continuum limit, where $\Delta x, \Delta t \longrightarrow 0$, 
and using Itô's Lemma \cite{book:gardiner},
one can show that the probability distribution satisfies
the functional Fokker-Planck-Kramers (FPK) equation of the scalar field: 
\begin{align}
&\partial_t \rho_t(\{\Pi, \phi\}) = \nonumber \\
& \int dk \left[ - \frac{\delta}{\delta \phi (-k)} 
\left(
\frac{\delta H(\{\Pi,\phi,b_t\})}{\delta \Pi(k)} 
- \gamma_\phi \frac{\delta H(\{\Pi,\phi,b_t\})}{\delta \phi(k)} 
- \frac{\gamma_\phi}{\beta} \frac{\delta}{\delta \phi (k)} 
\right)
\right.  \nonumber \\
&\left.  
\hspace{1.1cm} 
+ \frac{\delta}{\delta \Pi (-k)}
\left( 
\frac{\delta H(\{\Pi,\phi,b_t\})}{\delta \phi(k)}  
+ \gamma_\Pi \frac{\delta H(\{\Pi,\phi,b_t\})}{\delta \Pi(k)} 
+
\frac{\gamma_\Pi }{\beta} \frac{\delta}{\delta \Pi (k)}
\right)
\right]
 \rho_t(\{\Pi, \phi\}) 
 \, , \label{eqn:fpk}
\end{align}
See Appendix \ref{app:FPK} for details.
Here the functional derivatives and integrals are assumed to be defined in the continuum limit as
\begin{align}
\lim_{\Delta k \to 0}\frac{1}{\Delta k} \frac{\partial}{\partial \phi(k_n)} &=  \frac{\delta}{\delta \phi(k)} \, , 
\label{eqn:cont_lim1} \\ 
\lim_{\substack{\Delta k \to 0 \\ N \to \infty}} \Delta k \sum_{n=-N}^{N-1} &= \int^\infty_{-\infty} dk \, .\label{eqn:cont_lim2}
\end{align}
See Appendix \ref{app:continuum} for further discussions about this continuum limit.
In the absence of the interaction with the Brownian thermostat, 
the coefficients $\gamma_\phi $ and $ \gamma_\Pi $ vanish 
and thus the functional FPK equation is reduced to the Liouville equation for the field.

If the spectrum of the Hamiltonian $H$ is bounded from below and $b_t(k_n)$ becomes constant after a certain $t = \tau$, {\it i.e.},
$b_t(k_n) = b_\tau(k_n)$ for $t \ge \tau$, 
the functional FPK evolves an arbitrary initial condition towards the 
Gibbs state 
\begin{equation}\label{eq:equilstate}
\rho^\star (\{\Pi, \phi\}) = \frac{1}{Z(\tau)}e^{-\beta H(\{\Pi, \, \phi, \, b_{\tau}\})} \, ,
\end{equation} 
where the partition function is given by
\begin{equation}\label{eq:partfunc}
Z(\tau) = \int [\ud \Pi] \int [\ud \phi] e^{-\beta H(\{\Pi, \, \phi, \, b_{\tau}\})} \, .
\end{equation} 
See the discussion in Sec.\ \ref{sec:rel_ent} for details.

One may wonder why the stochastic differential equations are defined for $\widehat{\phi}_t (k_n)$ instead of $\widehat{\Phi}_t(x_i)$. 
It is because we need to introduce 
the coefficients $\gamma_\phi$ and $\gamma_\Pi$ 
depending on $k_n$. 
In our model, we can introduce heat, work and entropy which are consistent with thermodynamical interpretation 
for arbitrary real and positive functions $\gamma_\phi $ 
and $\gamma_\Pi $. 
However, for the corresponding quantum master equation, we use 
coefficients with a predetermined $k_n$ dependence, ensuring that the quantum master equation results in a CPTP evolution.


\section{Stochastic Thermodynamics for scalar fields} \label{sec:stother}

Let us first clarify our stance on incorporating a thermodynamical interpretation into our model. Our analysis of thermodynamical behavior excludes environmental data and relies solely on information obtained directly from the system itself. Consequently, in our framework, entropy pertains only to the system, allowing for the possibility of negative entropy change, unlike the total entropy change of the  
total system 
(system plus environment), which remains positive. Furthermore, we take a
standard view on the thermodynamical entropy, defining it as a quantity related to equilibrium states, i.e., a function only of macroscopic (thermodynamic) variables. In this paper, we demonstrate that Shannon's information entropy in stochastic thermodynamics and the von Neumann entropy in quantum thermodynamics adhere to laws analogous to the first and second laws of thermodynamics. 
Although these entropies can be calculated even for non-equilibrium states, 
they should not be indiscriminately equated with thermodynamical entropy itself.

\subsection{Introduction of classical heat}

In standard Brownian motion, 
it is assumed that the interaction with thermostat influences 
the motion only through the momentum equation, leaving the position equation unchanged. 
However, quantizing such standard Brownian motion 
yields a master equation that does not preserve complete positivity of the state \cite{oliveira1,oliveira2}. 
To address this, we must ensure that the position equation 
is modified by the interaction with thermostat, as discussed in Ref.\ \cite{KoiNic2023}. 
In other words, standard Brownian motion is recovered by setting $\gamma_\phi = 0$ in the model described by Eqs.\ (\ref{eqn:sde1}) and (\ref{eqn:sde2}). 
Although this is not a straightforward generalization, 
we can define heat, work, and entropy in a manner that satisfies laws analogous to those of thermodynamics, as we will demonstrate in this section.
Therefore, 
our generalized model provides a suitable phenomenological framework for describing 
thermal relaxation processes. This area has been the focus of intense debate in the theory 
of open quantum systems (see, for instance, \cite{dann, soret}).

In stochastic thermodynamics \cite{sekimoto}, the heat absorbed by the system from the Brownian thermostat is represented by the work done by the thermostat on the system \cite{sekimoto}.
To illustrate this definition, let us consider equations of motion for a particle given by
\begin{align}
\frac{\ud{\bf q}}{\ud t} &= \frac{\partial H}{\partial {\bf p}} + {\bf F}^{(q)}_{\rm ex} \, ,\\
\frac{\ud{\bf p}}{\ud t} &= - \frac{\partial H}{\partial {\bf q}} + {\bf F}^{(p)}_{\rm ex} \, ,
\end{align}
where $H$ is a particle Hamiltonian, and ${\bf F}^{(q)}_{\rm ex} $ and ${\bf F}^{(p)}_{\rm ex}$ are external perturbations to velocity and acceleration, respectively.
Then the change of the Hamiltonian is 
\begin{equation}\label{eq:ext_work}
\ud H({\bf q},{\bf p}) = {\bf F}^{(p)}_{\rm ex} \cdot \ud{\bf q} - {\bf F}^{(q)}_{\rm ex}\cdot \ud {\bf p}  \, ,
\end{equation}
that is, the right-hand side represents the work provided by the external perturbations. 
The standard definition of external work in classical mechanics is reproduced when ${\bf F}^{(q)}_{\rm ex} =0$.

We now apply this idea to define the heat for the scalar field system.
The interactions between the system and the Brownian thermostat are represented by the second and third terms on the right-hand sides of Eqs.\ (\ref{eqn:sde1}) and (\ref{eqn:sde2}), which, in the particle model, correspond to ``external perturbations'' in the above discussion.
Therefore the heat (work done by the external perturbations) will be defined by 
\begin{align} 
\ud \widehat{Q}_t 
&=
\Delta k \sum_{n=-N}^{N-1} \left( - \gamma_\Pi \frac{1}{\Delta k} \frac{\partial \widehat{H}(\{\widehat{\Pi}_t,\widehat{\phi}_t, b_t\})}{\partial \widehat{\Pi}_t (-k_n)}  + \sqrt{\frac{2\gamma_\Pi}{\Delta k \, \beta}} \frac{\ud \widehat{B}^{\Pi}_t (k_n)}{\ud t} \right)\circ \ud \widehat{\phi}_t(-k_n) \nonumber \\
&- \Delta k \sum_{n=-N}^{N-1} \left( - \gamma_\phi \frac{1}{\Delta k} \frac{\partial \widehat{H}(\{\widehat{\Pi}_t,\widehat{\phi}_t, b_t\})}{\partial \widehat{\phi}_t (-k_n)}  + 
\sqrt{\frac{2\gamma_\phi }{\Delta k \, \beta}} 
\frac{\ud \widehat{\cal B}^\phi_t(k_n)}{\ud t} \right)\circ \ud \widehat{\Pi}_t(-k_n)
\, , \label{eqn:st_heat}
\end{align}
where, instead of the scalar product in Eq.\ (\ref{eq:ext_work}),
the product between stochastic variables is given by the Stratonovich definition \cite{book:gardiner}:
\begin{equation}
\ud \widehat{B}^{\mu}_t(k_m) \circ 
f(\widehat{\Phi}_t(k_n)) 
:= \ud \widehat{B}^{\mu}_t(k_m) \frac{f(\widehat{\Phi}_t(k_n)) + f(\widehat{\Phi}_{t+\ud t}(k_n))}{2} \, , \label{eqn:strat_prod}
\end{equation}
for $\mu = \phi, \Pi$.

\subsection{First and second laws in stochastic thermodynamics}

It is natural to define the energy of the system by its Hamiltonian.
Moreover, as is done in Ref.\ \cite{sekimoto}, the work applied to the system is caused 
by the change of the external parameter $b_t(x_i)$ and thus defined by 
\begin{align}\label{eq:stowork}
\ud \widehat{W}_t &= \sum_{n=-N}^{N-1} \frac{\partial H(\{ \widehat{\Pi}_t,\widehat{\phi}_t,b_t \})}{\partial b_t( k_n)} \ud b_t( k_n)\, .
\end{align}
Then, for each stochastic event, we can write
\begin{align}
\ud \widehat{Q}_t 
&= 
\ud H(\{ \widehat{\Pi}_t,\widehat{\phi}_t, b_t\} )  
-  \ud \widehat{W}_t   \, , \label{eqn:1stlaw}
\end{align}
which corresponds to the first law in stochastic thermodynamics.
See, Appendix \ref{app:1st} for details.

Now, we consider the information entropy associated with the probability distribution $\rho$,
\begin{equation}\label{eqn:Sst}
S_{\rm ST} [\rho_t]= - k_{\rm B} \int [\ud \Pi] \int [\ud \phi] \rho_t(\{\Pi,\phi\}) \ln \rho_t(\{\Pi,\phi\}) \, , 
\end{equation}
and the the expectation value of the heat
\begin{equation}\label{eq:Eheat}
\lceil \ud \widehat{Q}_t   \rfloor :=
\int [\ud \Pi_0] \int [\ud \phi_0] \rho_0 (\{ \Pi_0, \phi_0 \}) \uE \left[ \ud \widehat{Q}_t \right] \, .
\end{equation}
Then, for arbitrary initial conditions and external protocols $b_t(k_n)$, 
we find that the following inequality is satisfied: 
\begin{align}
\lefteqn{\frac{\ud S_{\rm ST}[\rho_t]}{\ud t} - \frac{1}{T} \frac{1}{\ud t} \lceil \ud \widehat{Q}_t   \rfloor} & \nonumber \\
&=
k_{\rm B} 
\int [\ud \Pi] \int [\ud \phi]  \int \ud k   \frac{ \gamma_\phi \beta}{ \rho_t (\{\Pi, \phi\})}
\left|
\frac{\delta H}{\delta \phi(k)} \rho_t (\{\Pi,\phi\}) 
+
\frac{1}{\beta} \frac{\delta \rho_t (\{\Pi, \phi\})}{\delta \phi(k)} 
\right|^2 \nonumber \\
& + 
k_{\rm B} 
\int [\ud \Pi] \int [\ud \phi]  \int \ud k   \frac{ \gamma_\Pi\beta}{ \rho_t (\{\Pi, \phi\})}
\left|
\frac{\delta H}{\delta \Pi(k)} \rho_t (\{\Pi,\phi\}) 
+
\frac{1}{\beta} \frac{\delta \rho_t (\{\Pi, \phi\})}{\delta \Pi(k)} 
\right|^2 \ge 0 \, ,  
\label{eqn:st2nd}
\end{align}
which corresponds to the second law of thermodynamics. 
See Appendix \ref{app:2nd} for 
detailed calculations.
The equality is satisfied when $\rho = \rho^\star (\{ \Pi, \phi \})$ and 
$b_t(x_i) = b_\tau(x_i)$ for $t \ge \tau$, 
see also the discussion around Eqs.\ (\ref{eq:equilstate}) and (\ref{eq:partfunc}).

As was emphasized, the analogous laws introduced here are not equivalent to those in thermodynamics. 
For instance, the information entropy is defined for any non-equilibrium state, whereas thermodynamical entropy is 
a thermodynamical quantity only defined for equilibrium states. 
Moreover, we have not demonstrated that this entropy is expressed solely as a function of 
thermal equilibrium quantities.

Notably, in the above derivations, it is not necessary to specify the system Hamiltonian, allowing the first and second laws to be defined for any interacting scalar field. Consequently, stochastic thermodynamics is broadly applicable to various scalar field theories. However, this applicability becomes more limited when considering the quantum version of this model, as discussed in the following sections.

\section{Quantum Master Equation for Scalar Field} \label{sec:quather}

To provide a unified description of stochastic and quantum thermodynamics, 
we present a quantum master equation 
that reproduces the results in Sec.\ \ref{sec:stother} in its classical limit.
By extending the formulation proposed in Ref.\ \cite{KoiNic2024}, we derive such a field-theoretical quantum master equation 
by applying canonical quantization to the functional FPK equation (\ref{eqn:fpk}).

It is remarkable that the functional FPK equation (\ref{eqn:fpk})
can be expressed only  in terms of the Poisson brackets:
\begin{align}
\partial_t \rho_t (\{\Pi, \phi\})
&= 
-\{ \rho_t, H \}_{\rm PB}
+\int \ud k \,  \frac{\gamma_\phi }{\beta} \{ e^{-\beta H} \{ e^{\beta H} \rho_t, \Pi(k) \}_{\rm PB}, \Pi(-k) \}_{\rm PB} \nonumber \\
&+  \int \ud k\, \frac{\gamma_\Pi}{\beta} \{ e^{-\beta H} \{ e^{\beta H} \rho_t, \phi(k) \}_{\rm PB}, \phi(-k) \}_{\rm PB} \, ,
\end{align} \label{eqn:fpk2}
where
\begin{equation}
\{ f, g \}_{\rm PB}
= \int \ud k \left( 
\frac{\delta f}{\delta \phi(k)} \frac{\delta g}{\delta \Pi(-k)} 
-
\frac{\delta f}{\delta \Pi(-k)} \frac{\delta g}{\delta \phi(k)} 
\right) \, . \label{eqn:pb}
\end{equation}
The structure of this classical equation can be thus straightforwardly quantized.
Let us consider the following canonical quantization rules: 
\begin{align}
\begin{array}{ccc}
\rho_t (\{\Pi, \phi\}) & \longrightarrow & \hat{\rho}(t) \, ,\\
\{ f, g \}_{\rm PB} & \longrightarrow & -\frac{\ii}{\hbar} [\hat{f},\hat{g}] \, ,\\
e^{\pm \beta H} \rho_t (\{\Pi, \phi\})& \longrightarrow & e^{\pm \frac{\beta}{2} \hat{H}} \hat{\rho}(t) e^{\pm \frac{\beta}{2} \hat{H}} \, . \label{eqn:ccr}
\end{array}
\end{align}
The symbol ``$\, \hat{\bullet}\, $'' denotes operators and $\hat{\rho}(t)$ is the density matrix of the system. 
Using these replacements, our quantum master equation obtained from the functional FPK equation is 
\begin{align}
\partial_t \hat{\rho}(t)
&= 
\frac{\ii}{\hbar} \left[\rho(t), \hat{H}(t) \right]
- \int \ud k \, \frac{\gamma_\phi}{\beta \hbar^2} \left[ e^{-\frac{\beta}{2} \hat{H}(t)} \left[ e^{\frac{\beta}{2} \hat{H}(t)} \hat{\rho}(t) e^{\frac{\beta}{2} \hat{H}(t)}, \hat{\Pi}(k) \right] e^{-\frac{\beta}{2} \hat{H}(t)}, \hat{\Pi}(-k) \right] \nonumber \\
&-  \int \ud k \, \frac{\gamma_\Pi}{\beta \hbar^2} \left[ e^{-\frac{\beta}{2} \hat{H}(t)} \left[ e^{\frac{\beta}{2} \hat{H}(t)} \hat{\rho}(t) e^{\frac{\beta}{2} \hat{H}(t)}, \hat{\phi}(k) \right]e^{-\frac{\beta}{2} \hat{H}(t)}, \hat{\phi}(-k) \right] \label{eqn:qme}\, .
\end{align}
When there is no interaction with Brownian thermostat, {\it i.e.},
when both $\gamma_\phi =  \gamma_\phi (|k|,t)$ and $\gamma_\Pi = \gamma_\Pi(|k|,t)$ vanish, 
our quantum master equation is reduced to the Heisenberg equation of motion. 
Note that, if the spectrum of the Hamiltonian $\hat{H}(t) = H(\{ \hat{\Pi},\hat{\phi}, b_t \})$ is bounded from below and $b_t = b_{\tau}$ for $t \ge \tau$, 
the differential equation has a unique stationary solution (fixed-point attractor) in the asymptotic limit in time, which is given by Gibbs's thermal equilibrium state:
\begin{equation} \label{eq:eqstate}
\hat{\rho}^\star = \frac{1}{Z(\tau)}e^{-\beta \hat{H}(\tau)} \, ,
\end{equation} 
where 
\begin{equation}
Z(\tau) = {\rm Tr} [e^{-\beta \hat{H}(\tau)} ]\, .
\end{equation}

It should be however noted that, in general,
the evolution described by Eq.\ (\ref{eqn:qme}) does not satisfy
the requirements for a CPTP evolution.
To address this point,
we consider the simple case where the system Hamiltonian takes the form of a free scalar field with an explicit time dependence,
\begin{align}\label{eq:sysham}
\hat{H}(t)
= 
\int \ud k \left\{
\frac{c^2}{2} |\hat{\Pi} (k)|^2 + \frac{k^2}{2} |\hat{\phi} (k)|^2 + \frac{1}{2}b_t^2 (k) |\hat{\phi} (k)|^2
\right\}  \, ,
\end{align} 
introduced through the external parameter $b_t(k)$, which corresponds to the mass term.

For this Hamiltonian, Eq.\ (\ref{eqn:qme})
can be conveniently rewritten as
\begin{align}
\frac{\ud}{\ud t} \hat{\rho}(t) 
&= \frac{\ii}{\hbar} \left[\hat{\rho} (t), \hat{H}(t) \right] 
-  \frac{1}{\hbar}\int \ud k \, \hat{\rho}(t)
(\hat{\phi}(k),\hat{\Pi}(k)) {\bf L}^T (t)
\left(
\begin{smallmatrix}
\hat{\phi}(-k) \\
\hat{\Pi}(-k)
\end{smallmatrix}
\right)
\nonumber \\
&-  \frac{1}{\hbar}\int \ud k  \,
(\hat{\phi}(k),\hat{\Pi}(k)) {\bf L}^* (t)
\left(
\begin{smallmatrix}
\hat{\phi}(-k) \\
\hat{\Pi}(-k)
\end{smallmatrix}
\right)
\hat{\rho}(t)  \nonumber \\
&+ \frac{1}{\hbar} \int \ud k \,
(\hat{\phi}(k),\hat{\Pi}(k))
\hat{\rho}(t)
({\bf L} (t)+ {\bf L}^\dagger (t))
\left(
\begin{smallmatrix}
\hat{\phi}(-k) \\
\hat{\Pi}(-k)
\end{smallmatrix}
\right)
\, , \label{eqn:qmeq2}
\end{align}
where
\begin{align}
{\bf L}(t)
&= 
\hbar\left(
\begin{array}{cc}
\frac{\gamma_\Pi }{\beta \hbar^2} \cosh \Theta (t)& -\frac{\ii \gamma_\Pi c^2}{\beta \hbar^2 \omega_t (k)} \sinh \Theta (t) 
\\
 \frac{\ii \gamma_\phi  \omega_t (k)}{\beta \hbar^2 c^2} \sinh \Theta (t)
&  \frac{\gamma_\phi }{\beta \hbar^2} \cosh \Theta (t)
\end{array}
\right) \, ,
\end{align}
with
\begin{align}
\Theta(t) &=  \frac{\beta \hbar \omega_t (k)}{2} \, ,\\
\omega_{k}(t) &= \lim_{\Delta k \rightarrow 0} c \sqrt{\lambda_k^2 + b_t^2 (k)} =c \sqrt{k^2 + b_t^2 (k)}  \, .
\end{align}
See Eq.\ (\ref{eqn:lambda}) for the definition of $\lambda_k$.
As discussed in Ref.\ \cite{KoiNic2024}, 
the quantum master equation given by the form of Eq.\  (\ref{eqn:qmeq2}) describes a CPTP Markovian evolution 
when all eigenvalues of the matrix ${\bf L}_{\rm H} := {\bf L} + {\bf L}^\dagger$ are non-negative 
\cite{rivas2012}. 
Thus, 
since ${\bf L}_{\rm H}$ is a $2 \times 2$ matrix, their eigenvalues are non-negative 
if and only if $ {\rm det} {\bf L}_{\rm H} \ge 0$. 
Consequently,
the parameters $\gamma_\Pi $ and $\gamma_\phi$ should satisfy the following condition:
\begin{align}
\sinh^2 \Theta (t) \left\{ 4 \frac{\gamma_\Pi (|k|,t) \gamma_\phi (|k|,t)}{\beta^2 \hbar^4}\coth^2 \Theta (t)
-\left( 
 \frac{\gamma_\Pi (|k|,t) c^2}{\beta \hbar^2 \omega_t (k)}   +\frac{\gamma_\phi (|k|,t) \omega_t (k)}{\beta \hbar^2 c^2}  
\right)^2 
\right\} \ge 0 \, .
\label{eqn:detlh}
\end{align}
In the end, this inequality for any temperature (or for any $\Theta$) 
is always true if and only if
\begin{equation}
\gamma_\Pi (|k|,t)= \frac{\omega^2_t (k)}{c^4} \gamma_\phi (|k|,t)\, . \label{eqn:dbc}
\end{equation}
Although this relation depends on the wave number $k$, 
it is similar to the corresponding one in the quantum master equation for a harmonic oscillator. 
See the discussion below Eq.\ (35) in Ref.\ \cite{KoiNic2023}.

Using this relation, 
our quantum master equation is mapped into the well-known GKSL form:
\begin{align}\label{eq:LindME}
\frac{\ud}{\ud t}\hat{\rho}(t)
&=
\frac{\ii}{\hbar} [\rho(t), \hat{H} ] \nonumber \\
& -\frac{1}{2\hbar}   \int \ud k  
\sum_{i=\pm} \gamma_i (|k|,t)
\left[
 \hat{L}^\dagger_i (k) \hat{L}_i (k) \hat{\rho} (t) + \hat{\rho} (t)\hat{L}^\dagger_i (k) \hat{L}_i (k) 
-2  \hat{L}_i (k) \hat{\rho} (t) \hat{L}^\dagger_i (k)  
\right]
  \, ,
\end{align}
where Lindblad's jump operators are defined by 
\begin{align}
\hat{L}_+ (k)&= \hat{a}(k) \, ,\\
\hat{L}_- (k)&= \hat{a}^\dagger (k) \, , 
\end{align}
and we have introduced the 
Lindblad rates,    
\begin{align}\label{eq:param}
\gamma_\pm (|k|,t) 
:= 
\frac{4\gamma_\phi(|k|,t)}{\beta \hbar}\frac{\hbar \omega_t (k)}{c^2} e^{\pm \frac{\beta \hbar \omega_t (k)}{2}} \, . 
\end{align}
The creation-annihilation operators are defined by the transformation,
\begin{align}
\left( 
\begin{array}{c}
 \hat{a}^\dagger (-k) \\
\hat{a}(k)
\end{array}
\right)
= 
\sqrt{\frac{c^2}{2\hbar \omega_t (k)}} \left(
\begin{array}{cc}
1 & -\ii \\
1 & \ii
\end{array}
\right)
\left(
\begin{array}{c}
\frac{\omega_k(t)}{c^2} \hat{\phi}(-k) \\
\hat{\Pi} (k)
\end{array}
\right) \, ,
\end{align}
which diagonalizes the system Hamiltonian (\ref{eq:sysham}), 
\begin{equation}
\hat{H} (t) = \int \ud k \, \hbar \omega_t (k) \left( \hat{a}^\dagger (k)\hat{a} (k)+ \frac{1}{2} \right) \, .
\end{equation}
See Appendix \ref{app:continuum} for details about the continuum representation of 
the creation-annihilation operators.

Our quantization method is applicable regardless of whether the system under consideration follows Markovian or non-Markovian dynamics. 
In particular, as demonstrated in Ref.\ \cite{KoiNic2024}, when the coefficients $\gamma_i(k)$ are time-independent, 
a completely positive equation is always Markovian. 
More generally, if the coefficients are time-dependent, it is known that the positive semidefiniteness of ${\bf L}_{\rm H}$ is a necessary and sufficient condition for 
the system's time evolution to be Markovian and to maintain complete positivity \cite{rivas2012, breuer2016}. 
However, even if ${\bf L}_{\rm H}$ acquires negative eigenvalues at certain times, it is known that complete positivity may still be preserved in the non-Markovian evolution 
governed by Eq.\ (\ref{eq:LindME}). For example, the HPZ equation discussed in Ref.\ \cite{hpz}, which accounts for 
the non-equilibrium time evolution of the environmental variables, is non-Markovian where 
${\bf L}_{\rm H}$ can exhibit negative eigenvalues, 
yet it is considered to fulfill the CPTP condition.
We briefly discuss some aspects of this non-Markovianity in Sec.\ \ref{sec:concl}.

\section{Quantum thermodynamics}

In stochastic thermodynamics, the heat absorbed by the system is defined  by the work done by 
the Brownian thermostat, as shown in Sec.\ \ref{sec:stother}.
It is easy to confirm that the definition given by Eq.\ (\ref{eqn:st_heat}) is reexpressed in an alternative form, 
\begin{equation}\label{eq:meanstocheat}
\lceil \ud \widehat{Q}_t   \rfloor = \ud t \int [\ud \Pi] \int [\ud \phi] \frac{\partial \rho_t (\{\Pi,\phi\})}{\partial t} H  (\{ \Pi, \phi, b_t \}) \, , 
\end{equation}
where we used the continuum limit expressed in Eqs.\ (\ref{eqn:cont_lim1}) and (\ref{eqn:cont_lim2}), 
and the temporal derivative inside the integral is given by the functional FPK equation (\ref{eqn:fpk}).
Using the canonical quantization rules in Eq.\ (\ref{eqn:ccr}), 
the corresponding heat in quantum thermodynamics is defined by 
\begin{equation}\label{eqn:qheat}
\lceil \ud \widehat{Q}_t   \rfloor \longrightarrow 
\ud Q(t) = {\rm Tr} \left[ \ud t \frac{\ud \hat{\rho}(t)}{\ud t}  \hat{H} (t) \right]  \, .
\end{equation}
The work applied to the system is induced by the change of the external parameter 
$b_t (k)$ in the system Hamiltonian operator
and thus is defined by 
\begin{equation}
\ud W (t)=  {\rm Tr} \left[\hat{\rho} (t) \int \ud k\, \frac{\partial \hat{H} (t)}{\partial b_t(k)} \ud b_t(k) \right] \, . 
\end{equation}
Similar to the heat in Eq.\ (\ref{eq:meanstocheat}), the above expression can be derived from the canonical quantization 
of the mean stochastic work $\lceil \ud \widehat{W}_t \rfloor$ in Eq.\ (\ref{eq:stowork}) 
and the stochastic mean defined in Eq.\ (\ref{eq:Eheat}).

It is very natural to define the energy of the system by the expectation value of $\hat{H}(t)$, therefore the first law in quantum thermodynamics is represented by 
\begin{equation}
{\rm Tr}[\hat{\rho}(t+\ud t)\hat{H} (t+\ud t)] - {\rm Tr}[\hat{\rho}(t)\hat{H} (t)]
= \ud Q (t)+ \ud W (t) \, . \label{eqn:Quant_first_law}
\end{equation}
Comparing this with Eq.\ (\ref{eqn:1stlaw}), 
one can observe the quantum-classical correspondence in the law analogous to the first law of thermodynamics. 
Interestingly enough, the first law in quantum thermodynamics is derived from the quantization of classical stochastic motion, providing a clear interpretation of all three quantities in Eq.\ (\ref{eqn:Quant_first_law}). 
Despite being expressed exactly as Eq.\ (\ref{eqn:Quant_first_law}), 
the mean energy variation in the theory of quantum master equations is ambiguous 
and is an open problem \cite{NicMaia}.

Let us define the entropy of our quantum system by the von Neumann entropy,
\begin{equation}
S_{\rm QT} [\hat{\rho}(t)]= - k_{\rm B} {\rm Tr} \left[ \hat{\rho} (t)\ln \hat{\rho}(t) \right] \, . \label{eqn:vneu-ent}
\end{equation}
When the system Hamiltonian is given by Eq.\ (\ref{eq:sysham}), 
we can calculate the temporal derivative of $S_{\rm QT}$ using  
the GKSL equation (\ref{eq:LindME}) for an arbitrary initial state and an external parameter $b_t(k)$. 
We then obtain the following inequality involving $S_{\rm QT}$ and heat:
\begin{align}
\frac{\ud S_{\rm QT}[\hat{\rho}(t)]}{\ud t}- \frac{1}{T}\frac{\ud Q_t}{\ud t}
&=
k_{\rm B}\int \ud k \, \ud n \, \ud m \, 
 P_n (t)R^+_{mn} (k,t) \left\{ \ln \frac{P_n (t) R^{+} _{mn}(k,t)}{P_m (t) R^{-} _{nm}(k,t)} \right\}   \nonumber \\
& + k_{\rm B}\int \ud k \, \ud n \, \ud m \, 
 P_n (t) R^-_{mn} (k,t)\left\{ \ln \frac{P_n (t)  R^{-} _{mn}(k,t)}{P_m (t)  R^{+} _{nm}(k,t)}\right\}  \ge  0 \, , \label{eqn:qt2nd}
\end{align}
where $P_n(t)$ are the eigenvalues of the density operator associated to the eigenvectors $| k ,t\rangle$, that is,
\begin{equation}
\hat{\rho} (t) | k ,t\rangle = P_k (t) | k ,t\rangle \, ,
\end{equation}
and 
\begin{align}
R^{\pm} _{nm}(k,t) := \gamma_\pm (|k|,t) | \langle n,t | L_{\pm}(k) | m,t \rangle |^2 \, .
\end{align}

Assuming that the time-dependence of the system Hamiltonian vanishes after a certain time
 $\tau$, $b_t (k) = b_\tau (k)$ for $t\ge \tau$, 
our quantum master equation drives any state towards its stationary state, 
$\hat{\rho}^\star$, see Eq.(\ref{eq:eqstate}).
Further, the coefficients $\gamma_\pm (|k|,t)$, defined in Eq.\ (\ref{eq:param}), satisfies the detailed balance condition  
\begin{equation}
\frac{\gamma_- (|k|,t)}{\gamma_+ (|k|,t)} = 
\frac{\gamma_- (|k|,\tau)}{\gamma_+ (|k|,\tau)} = e^{-\beta \hbar \omega_\tau (k)} \, ,
\end{equation}
which means that the equality in Eq.\ (\ref{eqn:qt2nd}) holds for $\hat{\rho}^\star$.
This represents the second law in quantum thermodynamics, and the correspondence 
between the classical and quantum inequalities (\ref{eqn:st2nd}) and (\ref{eqn:qt2nd}) is evident. 
In this context, we can assert that a quantum-classical correspondence exists 
between the information entropy and the von Neumann entropy.

\section{Brief Remarks on the Relation Between Thermodynamical and Information-Theoretical Entropy} \label{sec:rel_ent}

As mentioned in the introduction, numerous studies have applied information-theoretical techniques 
to quantum field theory, often utilizing two entropies: 
the entanglement and the relative entropies 
\cite{nishioka,witten}. 
In our model, the density matrix 
$\hat{\rho}$ does not describe the state of the environment and is therefore a reduced density 
matrix. 
Consequently, the von Neumann entropy introduced in Eq.\ (\ref{eqn:vneu-ent}) is 
identified with the entanglement entropy, 
which serves as a measure for quantitatively evaluating the degree of entanglement in quantum systems.
Relative entropy satisfies a specific inequality and is sometimes associated with the second law of thermodynamics. 
However, it is generally not equivalent to the entropy used in quantum thermodynamics.

Before discussing relative entropy, we first investigate the Kullback-Leibler divergence, which is 
extended to relative entropy in quantum theory.
In our field-theoretical model interacting with Brownian thermostats, 
it is defined by
\begin{align}
S_{\rm KL} (\rho_t| \rho^{\star}_t) &= \int [\ud \Pi] \int [\ud \phi]\,  \rho_t (\{\Pi, \phi\}) \ln \rho_t (\{\Pi, \phi\})  \nonumber \\
& -  \int [\ud \Pi] \int [\ud \phi]\,  \rho_t (\{\Pi, \phi\}) \ln \rho^{\star}_t (\{\Pi, \phi\})
\, ,
\end{align}
where $\rho^{\star}_t (\{\Pi, \phi\})$ is the time-local stationary solution of the functional FPK equation,
\begin{equation}
\rho^{\star}_t (\{\Pi, \phi\}) = \frac{1}{Z(t)} e^{-\beta H(t)} \, ,
\end{equation}
with $Z(t) = \int [\ud \Pi] \int [\ud \phi]\, e^{-\beta H(t)}$.
It can be shown that the time derivative is given by
\begin{align}
\frac{\ud S_{\rm KL}  (\rho_t| \rho^{\star}_t)}{\ud t} 
= 
&-\int [\ud \Pi] \int [\ud \phi]\, 
 \int \ud k \, 
 \frac{\gamma_\phi}{\beta} \rho_t \left| \frac{\delta \ln \rho_t}{\delta \phi(k)} -  \frac{\delta \ln \rho^{\star}_t}{\delta \phi(k)} \right|^2 \nonumber \\
&- \int [\ud \Pi] \int [\ud \phi]\, 
\int \ud k \,
\frac{\gamma_\Pi}{\beta} \rho_t
\left| \frac{\delta \ln \rho_t}{\delta \Pi(k)} -  \frac{\delta \ln \rho^{\star}_t}{\delta \Pi(k)} \right|^2 \nonumber \\
&
- \int [\ud \Pi] \int [\ud \phi]\, \rho_t \partial_t \ln \rho^{\star}_t  \, . \label{eqn:monotonicity}
\end{align}
Considering the informational entropy defined in Eq.\ (\ref{eqn:Sst}) and using the ``stochastic second law'', expressed in Eq.\ (\ref{eqn:st2nd}), we have 
\begin{align}
\frac{\ud \tilde{S}_{\rm KL} (\rho_t| \rho^{\star}_t)}{\ud t}
& = \frac{1}{k_B} \left\{
\frac{\ud S_{\rm ST}[\rho_t]}{\ud t} - \frac{1}{T} \frac{1}{\ud t} \lceil \ud \widehat{Q}_t   \rfloor
\right\} \ge 0 \, ,
\end{align}
where
\begin{equation}
\tilde{S}_{\rm KL}  (\rho_t| \rho^{\star}_t) := - S_{\rm KL} (\rho_t| \rho^{\star}_t)- \int^t \ud s \, \int [\ud \Pi] \int [\ud \phi]\, \rho_s \partial_s \ln \rho^{\star}_s \, . \label{eqn:mod_KL}
\end{equation}
Therefore, the time derivative of the Kullback–Leibler divergence is distinct from the second law of stochastic thermodynamics, because of 
the rightmost term in Eq.\ (\ref{eqn:mod_KL}).
When the Hamiltonian is independent of time, $H(t) = H$, 
the third line on the right-hand side of Eq.\ (\ref{eqn:monotonicity}) vanishes and 
the Kullback-Leibler divergence serves as a Lyapunov function (or H function), since the probability distribution $\rho$ evolves 
monotonically toward the stationary state $\rho^\star$ for any initial distribution. 
In this case,
the time derivative of the Kullback-Leibler divergence (\ref{eqn:monotonicity}) reduces to the second law in Eq.\ (\ref{eqn:st2nd}).

We now turn to the quantum generalization of the Kullback-Leibler divergence, the relative entropy, which is defined by 
 \begin{equation}
S_{\rm rel} (\hat{\rho}(t) | \hat{\rho}^\star (t)) = {\rm Tr} [ \hat{\rho}(t) \ln \hat{\rho}(t) ] - {\rm Tr} [\hat{\rho}(t) \ln \hat{\rho}^{\star} (t)] \, ,
\end{equation}
where
\begin{equation}
\hat{\rho}^{\star} (t) := \frac{1}{Z(t)} e^{-\beta \hat{H}(t)} \, ,
\end{equation}
with $Z(t) ={\rm Tr}[e^{-\beta \hat{H}(t)}  ]$.
Using the expression for heat in Eq.~(\ref{eqn:qheat}), 
it is possible to write (see Refs.\ \cite{deffner,breuer22}, for instance)
\begin{equation}
\frac{\ud S_{\rm QT}[\hat{\rho}(t)]}{\ud t}- \frac{1}{T}\frac{\ud Q_t}{\ud t}
=\frac{\ud}{\ud t}\tilde{S}_{\rm rel} (\hat{\rho}(t) | \hat{\rho}^\star (t)) \, ,
\end{equation}
where
\begin{equation}\label{eqn:tildeSrel}
\tilde{S}_{\rm rel} (\hat{\rho}(t) | \hat{\rho}^\star (t)) = - S_{\rm rel} (\hat{\rho} (t)| \hat{\rho}^{\star} (t)) - \int^t \ud s \, {\rm Tr} [\hat{\rho} (s)\partial_s \ln \hat{\rho}^{\star} (s)] \, .
\end{equation}
This modified relative entropy corresponds to the classical relation in Eq.\ (\ref{eqn:mod_KL}), leading to 
\begin{equation}
\frac{\ud}{\ud t}\tilde{S}_{\rm rel} (\hat{\rho}(t) | \hat{\rho}^\star (t)) \ge 0 \, ,
\end{equation}
with the help of Eq.\ (\ref{eqn:qt2nd}).
Once again, the time derivative of the relative entropy is distinct from the second law of quantum thermodynamics. 
The difference arises from the rightmost term in Eq.\ (\ref{eqn:tildeSrel}).


\section{Concluding remarks} \label{sec:concl}

In this paper, we studied a systematic procedure for deriving a quantum master equation which describes thermal relaxation processes in the scalar field theory.
Because it is known that the classical limit of open quantum dynamics satisfying the CPTP map does not coincide with standard Brownian motion, 
we first introduced a generalized field-theoretical model interacting with Brownian thermostat for the scalar field and confirmed that the model is consistent with stochastic thermodynamics. 
We then applied canonical quantization to this model to derive a quantum master equation.
While this equation is applicable to all forms of the scalar field Hamiltonian, it generally describes non-CPTP evolution. 
However, at least when the system Hamiltonian is given by a free scalar field, we demonstrated that by adjusting the parameters within our model, 
the quantum master equation gives the GKSL form.
In this framework, heat, work, and entropy are defined in such a way that they satisfy analogs of the first and second laws of thermodynamics. 
Since our quantum master equation is derived through canonical quantization applied to a model consistent with stochastic thermodynamics, 
the classical limit of quantum thermodynamics recovers stochastic thermodynamics in cases where quantum fluctuations vanish in the macroscopic limit \cite{KoiNic2024}.

This result suggests that the quantum-classical correspondence extends beyond closed systems governed by unitary time evolution, 
encompassing open systems.
Thus it is interesting to study Onsager's regression hypothesis where the average regression 
of thermal fluctuations behaves like the corresponding macroscopic irreversible process.
For quantum systems, it is considered that this hypothesis is violated \cite{grabert,talkner,ford,guarnieri,cosaccchi} and, consequently, 
it is worth investigating this property in the perspective of quantum-classical correspondence 
in open systems.

Our classical model takes into account the fact that the conventional linear relation between momentum and velocity does not necessarily hold due to dissipative effects. At first glance, this approach might seem to deviate from the conventional framework of classical physics; however, as exemplified by phenomena such as the Lorentz force, when forces depend on velocity, the simple linear relation between momentum and velocity generally fails. In other words, as a model that reflects the velocity dependence of forces in dissipative systems, the extension introduced in our model is a natural outcome.
Furthermore, as demonstrated by our model, maintaining the CPTP condition requires strict adherence to the criterion expressed in Eq.\ (\ref{eqn:dbc}). If one attempts to recover the conventional linear momentum–velocity relation by setting $\gamma_\phi = 0$, one is forced to also set $\gamma_\Pi = 0$, thereby eliminating any possibility of describing thermal relaxation phenomena. Indeed, every known quantum model that captures thermal relaxation processes incorporates a term corresponding to $\gamma_\phi$.

In quantum thermodynamics, we define entropy as the von Neumann entropy of the reduced density matrix obtained by coarse graining over the bath degrees of freedom. This coincides with the entanglement entropy used in information-theoretical treatments of quantum field theory. Another key quantity is the relative entropy, which serves as a natural quantum analogue of thermodynamical entropy because of its monotonicity. However, as noted in Refs.\ \cite{deffner,breuer22}, the time derivative of the relative entropy can be identified with the second law of quantum thermodynamics only when the system Hamiltonian is time-independent. The same caveat applies in stochastic thermodynamics: the time derivative of the Kullback–Leibler divergence corresponds to the second law only for a time independent Hamiltonian.

Our previous findings in Ref.\ \cite{KoiNic2024} indicate that the classical limit of our quantum master equation invariably yields the classical FPK equation (\ref{eqn:fpk}). 
This holds true irrespective of any non-Markovian or non-completely positive character of the underlying quantum equation. 
Moreover, in the classical limit, the eigenvalues of ${\bf L}_{\rm H}$ are always positive semidefinite, regardless of the parameter values.
Consequently, even if a non-Markovian quantum master equation were to possess Eqs.\ (\ref{eqn:sde1}) and (\ref{eqn:sde2}), 
and hence (\ref{eqn:fpk}) as its classical limit, any associated memory effects must vanish in this limit.
To investigate more the difference between non-Markovian dynamics and ours,  
we consider the HPZ equation given by Eqs.\ (17) and (18) in Ref.\ \cite{ferialdi}, 
which can be compared with Eqs.\ (28)-(35) in our paper \cite{KoiNic2023}. 
The dissipative term in our  Eq.\ (31) is expressed in terms of four jump operators, 
with three of its eigenvalues positive and one negative. 
In contrast, the corresponding term in the HPZ equation is written with two operators, with one negative eigenvalue. 
Furthermore, if we apply a straightforward Markov approximation (i.e., without modifications to ensure CP) to the HPZ equation, 
the HPZ equation reduces to the Joos-Zeh type equation given by Eq.\ (11) in Ref.\ \cite{ferialdi}. 
By contrast, it is not possible to recover the Joos-Zeh equation by merely adjusting the parameters of our equation.
Due to these structural differences, it is unlikely that our equation will recover complete positivity solely by incorporating memory effects through time-dependent parameters, whereas the HPZ equation maintains complete positivity.

In this paper, we studied general aspects of stochastic and quantum thermodynamics, but, we did not specifically examine how to evaluate 
heat, work, and entropy within a particular model. This task is far from trivial. For instance, in the applications to quantum field theory,
 it is known that the von Neumann entropy suffers from ultraviolet divergences \cite{witten}.
To the best of our knowledge, no systematic procedure for regularizing and renormalizing these divergences has been established. 
Note however that the first and second laws of quantum thermodynamics are expressed in terms of changes in such quantities. Therefore, if these divergences are regularized 
by subtracting constant terms, these laws will be applied without modification, even in the presence of such divergences.

Applying this method to the complex scalar field theory, we can construct a field-theoretical model with a conserved charge. 
However, it remains an open question whether such a model satisfies analogs of the first and second laws of thermodynamics, incorporating contributions from a chemical potential. It is important to comparing its behavior with that of the particle system proposed in Ref.\ \cite{sekimoto2000}.
As an alternative approach, see also Ref.\ \cite{carsten} as an example of field-theoretical systems with chemical potential.

Dissipative dynamics in the scalar field theory has been studied in cosmology and elementary particle physics. 
See, for example, Refs.\ \cite{morikawa,ramos1,ramos2,lombardo,greiner,rischke,biro,boya1,boya2,cooper,randrup,koide2002,wesp} 
and references therein.
In these approaches however, the main concern is the time evolutions of the field expectation values, such as order parameters in phase transitions, and 
the violation of the CPTP condition has not received much attention.
It is thus interesting to ask how the dynamics of phase transitions is affected by the requirement of complete positivity. 
The model presented in this paper 
provide a tool to study this aspect in non-equilibrium dynamics in field theory.

So far, our method is applicable only to bosonic or fermionic systems with quadratic Hamiltonians.
Nevertheless, it is known that even such relatively simple quadratic models can exhibit phase transitions 
as exemplified by the Su-Schrieffer-Heeger model, the Kitaev chain, and the XX spin model.
We believe that applying quantum thermodynamic analysis to such systems would be highly meaningful,
and we regard this as a promising direction for future research.

In Ref.\ \cite{collins}, the scalar field is decomposed into long and short wavelength components, 
and the coarse-grained dynamics is obtained by integrating out the short wavelength part. 
This type of coarse-graining is well-known and has been studied, for instance, 
using the influence functional method \cite{greiner,hpz} and the projection operator method \cite{koide2002}.
However, unlike these approaches, Ref.\ \cite{collins} focuses on the dynamics of the diagonal part of the density matrix, that is assumed to be governed by the functional Fokker-Planck equation, which is different from the functional FPK equation in the present paper. 
Specifically, the probability density of the scalar field configuration is described in a space spanned by the scalar field $\Phi (x)$.
This is different from our formulation of quantum thermodynamics, where the probability density is defined in phase space, spanned by both the scalar field 
$\Phi (x)$ and its conjugate field $\pi(x)$. 
The advantage of Ref.\ \cite{collins} is that the parameters in the functional Fokker-Planck equation can be derived perturbatively 
from microscopic dynamics. Their method might be extended to determine the parameters in our quantum master equation.

\begin{acknowledgments}

T.K. thanks the fruitful discussion with theory groups of the 
Institute for Theoretical Physics in the Johann Wolfgang Goethe University 
and the Frankfurt Institute for Advanced Studies (FIAS).
T.K. acknowledges the financial support by CNPq (No.\ 305654/2021-7) 
and the Fueck-Stiftung.
A part of this work has been done under the project INCT-Nuclear Physics
and Applications (No.\ 464898/2014-5);
F.N. is a member of the Brazilian National Institute of Science and Technology
for Quantum Information [CNPq INCT-IQ (465469/2014-0)].
\end{acknowledgments}

\appendix

\section{Derivation of functional FPK equation  -- Eq.(\ref{eqn:fpk}) } \label{app:FPK}

To calculate $\ud \rho_t (\{ \Pi, \phi \})$, we will take the temporal derivative of the
field configuration in Eq.\ (\ref{eqn:field_conf}) and use the correlations
(\ref{eqn:correlations_dbk}). 
Using Ito's lemma, a kind of Taylor expansion for stochastic variables \cite{book:gardiner}, we find
\begin{align} 
\ud\rho_t &=   \sum_{n=-N}^{N-1} \left[ 
\ud \widehat{\Pi}_t (k_n)  \frac{\partial}{\partial \widehat{\Pi}_t (k_n)} 
+ \ud \widehat{\phi}_t (k_n) \frac{\partial}{\partial \widehat{\phi}_t (k_n)} \right] 
\rho_t  \nonumber \\
& + \frac{1}{2} \sum_{n,m=-N}^{N-1} \left[
\ud \widehat{\phi}_t (k_n) \frac{\partial}{\partial \widehat{\phi}_t (k_n)} 
\ud \widehat{\phi}_t (k_m) \frac{\partial}{\partial \widehat{\phi}_t (k_m)} +
\ud \widehat{\Pi}_t (k_n) \frac{\partial}{\partial \widehat{\Pi}_t (k_n)} 
\ud \widehat{\Pi}_t (k_m) \frac{\partial}{\partial \widehat{\Pi}_t (k_m)} \right] \rho_t \nonumber \\
&+ O( (\ud t)^{3/2} )
\, .
\end{align}
Inserting Eqs.\  (\ref{eqn:sde1}), (\ref{eqn:sde2}) and (\ref{eqn:field_conf}) into, respectively, $\ud \widehat{\phi}_t$, $\ud \widehat{\Pi}_t$ and $\rho_t$, 
in the above equation and using the correlation properties 
in Eq.\ (\ref{eqn:correlations_dbk}), we find  
\begin{align}
\ud \rho_t  & = - \ud t \sum_{n=-N}^{N-1}
 \frac{\partial}{\partial \Pi (k_n)}
\left( 
- \frac{1}{\Delta k} \frac{\partial H}{\partial \phi(-k_n)}  
- \frac{\gamma_\Pi }{\Delta k} \frac{\partial H}{\partial \Pi(-k_n)} 
\right)
\rho_t  \nonumber \\
& - \ud t \sum_{n=-N}^{N-1}\frac{\partial}{\partial \phi (k_n)} 
\left(
 \frac{1}{\Delta k} \frac{\partial H}{\partial \Pi(-k_n)} 
- \frac{\gamma_\phi}{\Delta k} \frac{\partial H}{\partial \phi(-k_n)} 
\right)
\rho_t \nonumber \\
& +  \ud t \sum_{n=-N}^{N-1} 
\left(  \frac{\gamma_\phi}{\Delta k \beta} \frac{\partial}{\partial \phi (k_n)} 
\frac{\partial}{\partial \phi (-k_n)} 
+\frac{\gamma_\Pi}{\Delta k \beta} 
\frac{\partial}{\partial \Pi (k_n)}\frac{\partial}{\partial \Pi (-k_n)} \right)
\rho_t   + O( (\ud t)^{3/2} ) \, .
\end{align}
It is easy to see that this becomes the functional FPK equation in Eq.\ (\ref{eqn:fpk}) 
in the continuum limit. 

\section{Continuum limit} \label{app:continuum}

The continuum limit refers to consider the following limits:
\begin{align}
\Delta x\, , \Delta k \longrightarrow 0 \, ,\,\,\,
N \longrightarrow \infty \, .
\end{align}
After taking these limits, 
we further consider $\ell \longrightarrow \infty$ 
in the end of the calculations. 
The continuum version for the eigenvectors of the Laplacian in Eq.\ (\ref{eqn:eigvec}) behave as 
\begin{equation}
\tilde{u}^{(k)} (x)
:= \lim_{\Delta k \rightarrow 0} \frac{1}{\sqrt{\Delta k}} {u}^{(k_n)} (x)  = 
\frac{1}{\sqrt{2\pi}} e^{\ii k x} \, .
\end{equation}
The orthogonal condition in Eq.\ (\ref{eqn:orthocond}) 
and complete set condition in Eq.\ (\ref{eqn:completeset}) are, respectively, given by  
\begin{align}
&\int^\infty_{-\infty} \ud k \,  (\tilde{u}^{(k)} (x) )^\dagger \tilde{u}^{(k)} (x^\prime) = 
\delta(x - x^\prime)  \, , \\
&\int^\infty_{-\infty}  \ud x\, (\tilde{u}^{(k)}(x))^\dagger \tilde{u}^{(k^\prime)}(x) = 
\delta (k-k^\prime) \, . 
\end{align}

Then the expansions of the scalar field and the conjugate field, 
Eq.\ (\ref{eqn:Phi_phi}) and Eq. (\ref{eqn:pi_Pi}), 
are reduced to the well-known representations: 
\begin{align}
{\Phi} (x_i) = \sqrt{\Delta k} \sum_{n=-N}^{N-1} {\phi}({k_n}) u^{(k_n)} (x_i)
\xrightarrow{\rm continuum }
\Phi(x) = \int \ud k \, \phi(k) \frac{1}{\sqrt{2\pi}}e^{\ii kx} \, , \label{eqn:cl_Phi}\\
\pi (x_i) = \sqrt{\Delta k} \sum_{n=-N}^{N-1} {\Pi}(k_n) u^{(k_n)} (x_i)
\xrightarrow{\rm continuum }
\pi (x) = \int \ud k \, \Pi(k) \frac{1}{\sqrt{2\pi}}e^{\ii kx} \, . \label{eqn:cl_pi}
\end{align} 

From the quantization rules in Eq.\ (\ref{eqn:ccr}), 
we promote the coefficients in above expansion to the field operators 
$\hat \phi(k_n)$ and $\hat{\Pi}(k_n)$, which can be written in terms of (discrete) creation-annihilation operators:
\begin{align}
 \hat{\phi} ({k_n}) 
&= \sqrt{\frac{\hbar}{2\omega_{k_n} \Delta k}}
{ \, c \, }
(\hat{a}_{k_n} + \hat{a}^\dagger_{-k_n}) \, ,\\
\hat{\Pi}({k_n})
&= - \frac{\ii}{c} \sqrt{\frac{\hbar \omega_{k_n}}{2\Delta k}} (\hat{a}_{k_n} - \hat{a}^\dagger_{-k_n}) \, .
\end{align}

The discrete version of the creation-annihilation operators  satisfy 
\begin{equation}
[\hat{a}_{k_n}, \hat{a}^\dagger_{k_m}] = \delta_{k_n,k_m} \,  , 
\end{equation}
and, in the continuum limit, become 
\begin{equation}
[\hat{a}(k), \hat{a}^\dagger (k^\prime)] = \delta (k-k^\prime) \,  .
\end{equation}

The above definitions are introduced so that 
the canonical quantization rule in the $x$-representation is automatically satisfied in the continuum limit, 
\begin{equation}
[\hat{\Phi}_t (x), \hat{\pi}_t (x^\prime)] = \ii \hbar \delta(x-x^\prime) \, .
\end{equation}

\section{Derivation of first law in stochastic thermodynamics -- Eq.\ (\ref{eqn:1stlaw})} \label{app:1st}

In the following derivation, 
the Stratonovich product in Eq.\ (\ref{eqn:strat_prod}) will be implemented using a Taylor expansion:
\begin{equation}
\ud \widehat{B}^{\mu}_t(x_i) \circ f(\widehat{\Phi}_t(x_i)) = 
\ud \widehat{B}^{\mu}_t(x_i) f(\widehat{\Phi}_t(x_i)) + 
\frac{1}{2} \ud \widehat{B}^{\mu}_t(x_i) \frac{\partial f}{\partial \widehat{\Phi}_t(x_i)}
\frac{\partial \widehat{\Phi}_t(x_i)}{\partial t} \ud t + O((\ud t)^2) \, . 
\end{equation}
Note that this product is symmetric and distributive, 
$A \circ B = B \circ A$ and $A \circ (B+C) = A \circ B + A \circ C$. 
As an example of a term appearing in Eq.\ (\ref{eqn:st_heat}), we have
\begin{equation}
\begin{aligned}
\frac{\partial \widehat{H}}{\partial \widehat{\Pi}_t (-k_n)} \circ 
\ud \widehat{B}_t^\phi(k_n) & = 
\frac{\partial \widehat{H}}{\partial \widehat{\Pi}_t (-k_n)} \ud \widehat{B}_t^\phi(k_n) \nonumber \\
& + \frac{1}{2} \sum_m \frac{ \partial^2 \widehat{H} }
                 { \partial \widehat{\phi}_t (k_m) \partial \widehat{\Pi}_t (-k_n) } 
            \ud \widehat{\phi}_t(k_m) \, \ud \widehat{B}_t^\phi(k_n)  
 + O((\ud t)^2)  \, .        
\end{aligned}
\end{equation}
Using this result in the definition of heat given by Eq.\ (\ref{eqn:st_heat}),
we find  
\begin{align}
\ud \widehat{Q}_t & = 
\frac{\partial \widehat{H}}{\partial \widehat{\Pi}_t (k_n)} \, \ud t \, \circ
\sum_{n=-N}^{N-1} \left( -  \frac{\gamma_\Pi}{\Delta k} \frac{\partial \widehat{H}}{\partial \widehat{\Pi}_t (-k_n)} 
+ \sqrt{\frac{2\gamma_\Pi}{\Delta k \beta}} \frac{\ud \widehat{\cal B}^{\Pi}_t (k_n)}{\ud t} \right)  \nonumber \\
& + \frac{\partial \widehat{H}}{\partial \widehat{\phi}_t (k_n) } \,  \ud t \circ  
\sum_{n=-N}^{N-1} \left( - \frac{\gamma_\phi}{\Delta k} \frac{\partial \widehat{H}}{\partial \widehat{\phi}_t (-k_n)} + \sqrt{\frac{2\gamma_\phi}{\Delta k \beta}} \frac{\ud \widehat{\cal B}^\phi_t (k_n)}{\ud t}\right) 
+ O((\ud t)^{3/2})
\, . \label{eqn:st_heat_ap_1}
\end{align}
Adding and subtracting the term 
\begin{equation}
\sum_{n=-N}^{N-1}  \frac{\ud t}{\Delta k} 
\frac{\partial \widehat{H}}{\partial \widehat{\phi}_t (-k_n)} 
\frac{\partial \widehat{H}}{\partial \widehat{\Pi}_t (k_n)} 
=  \sum_{n=-N}^{N-1} \frac{\ud t}{\Delta k} 
\frac{\partial \widehat{H}}{\partial \widehat{\phi}_t (-k_n)} \circ
\frac{\partial \widehat{H}}{\partial \widehat{\Pi}_t (k_n)} \, ,
\end{equation}
on the right-hand side of Eq.\ (\ref{eqn:st_heat_ap_1}), this expression in the infinitesimal limit of $\ud t$ becomes
\begin{align}
\ud \widehat{Q}_t & =    
\frac{\partial \widehat{H}}{\partial \widehat{\Pi}_t (k_n)} \circ \!
\sum_{n=-N}^{N-1} \!\!\left( \!
- \frac{1}{\Delta k} \frac{\partial \widehat{H}}{\partial \widehat{\phi}_t (-k_n)}  
- \frac{\gamma_\Pi}{\Delta k} \frac{\partial \widehat{H}}{\partial \widehat{\Pi}_t (-k_n)} 
+ \sqrt{\frac{2\gamma_\Pi }{\Delta k \beta}} \frac{\ud \widehat{\cal B}^{\Pi}_t (k_n)}{\ud t} \!\right) \! \ud t \nonumber\\
& + \frac{\partial \widehat{H}}{\partial \widehat{\phi}_t (k_n)} \circ 
\sum_{n=-N}^{N-1} \left( 
\frac{1}{\Delta k} \frac{\partial \widehat{H}}{\partial \widehat{\Pi}_t (-k_n)}  
-  \frac{\gamma_\phi}{\Delta k} \frac{\partial \widehat{H}}{\partial \widehat{\phi}_t (-k_n)} + \sqrt{\frac{2\gamma_\phi }{\Delta k \beta}} \frac{\ud \widehat{\cal B}^\phi_t (k_n)}{\ud t}\right) \ud t \nonumber \\
&= 
\sum_{n=-N}^{N-1} \left( 
\frac{\partial \widehat{H}}{\partial \widehat{\phi}_t (k_n)} \circ \ud \widehat{\phi}_t (k_n) 
+
\frac{\partial \widehat{H}}{\partial \widehat{\Pi}_t (k_n)} \circ \ud \widehat{\Pi}_t (k_n)  
\right) \ . 
\end{align}
This equals to  Eq.\ (\ref{eqn:1stlaw}) and can be written as 
$\ud H(\{ \widehat{\Pi}_t,\widehat{\phi}_t,b_t \}) -  \ud \widehat{W}_t$, 
considering the definition in (\ref{eq:stowork}) for $\ud \widehat{W}_t$.

\section{Derivation of second law in stochastic thermodynamics -- Eq.(\ref{eqn:st2nd})} \label{app:2nd}

Using the functional FPK equation (\ref{eqn:fpk}) and integration by parts, 
the derivative of $S_{\rm ST} [\rho_t]$ is calculated as 
\begin{align}
& \frac{\ud S_{\rm ST}[\rho_t]}{\ud t} = 
- k_B \int [\ud \Pi] \int [\ud \phi] \, \partial_t \rho_t (\{\Pi,\phi\}) \,  
\ln \rho_t (\{\Pi,\phi\})  \nonumber \\
& = k_B \int [\ud \Pi] \int [\ud \phi]  \int \ud k 
\left[ 
\gamma_\phi \frac{\delta H}{\delta \phi(k)}  
\frac{\delta \rho_t }{\delta \phi(-k)} + 
\frac{\gamma_\phi}{\beta} \frac{1}{\rho_t} 
\left| \frac{\delta \rho_t }{\delta \phi(k)} \right|^2
  \right]  \nonumber \\
& + k_B \int [\ud \Pi] \int [\ud \phi]  \int \ud k 
\left[ 
\gamma_\Pi \frac{\delta H}{\delta \Pi(k)}  
\frac{\delta \rho_t }{\delta \Pi(-k)} 
+ \frac{\gamma_\Pi}{\beta} \frac{1}{\rho_t } \left| \frac{\delta \rho_t}{\delta \Pi(k)} \right|^2
  \right]  \, ,  
\end{align}
where we used the Poisson bracket, see Eq.(\ref{eqn:pb}),  
\begin{equation}
 \int [\ud \Pi] \int [\ud \phi] \{  H(\{\Pi, \phi\}), \rho_t(\{\Pi, \phi\})\}_{\rm PB} 
 = 0 \, .  
\end{equation}

Now, let us calculate the expectation value of heat given by Eq.\ (\ref{eqn:st_heat}). 
To this end, we will depart from the expression obtained in Eq.\ (\ref{eqn:st_heat_ap_1}):
\begin{align}
\uE [\ud \widehat{Q}_t ] & = 
\uE \left[ \frac{\partial \widehat{H}}{\partial \widehat{\Pi}_t (k_n)} \, \ud t \, \circ
\sum_{n=-N}^{N-1} \left( -  \frac{\gamma_\Pi}{\Delta k} \frac{\partial \widehat{H}}{\partial \widehat{\Pi}_t (-k_n)} 
+ \sqrt{\frac{2\gamma_\Pi}{\Delta k \beta}} \frac{\ud \widehat{\cal B}^{\Pi}_t (k_n)}{\ud t} \right) +  \right. \nonumber \\
& \hspace{1.1cm} \left. \frac{\partial \widehat{H}}{\partial \widehat{\phi}_t (k_n) } \,  \ud t \circ  
\sum_{n=-N}^{N-1} \left( - \frac{\gamma_\phi}{\Delta k} \frac{\partial \widehat{H}}{\partial \widehat{\phi}_t (-k_n)} + \sqrt{\frac{2\gamma_\phi}{\Delta k \beta}} \frac{\ud \widehat{\cal B}^\phi_t (k_n)}{\ud t}\right) \, \right] \nonumber \\
& = \uE \left[
(\Delta k)(\ud t) \sum_{n=-N}^{N-1} 
\left( - \frac{\gamma_\Pi}{(\Delta k)^2}\left| 
\frac{\partial \widehat{H}}{\partial \widehat{\Pi}_t (k_n)} \right|^2 
+ \frac{\gamma_\Pi }{(\Delta k)^2 \beta} 
\frac{\partial^2 \widehat{H}}{\partial \widehat{\Pi}_t (k_n) \partial \widehat{\Pi}_t (-k_n)}
\right) + \right. 
\nonumber \\
& \hspace{1.cm} \left.  (\Delta k) (\ud t)\sum_{n=-N}^{N-1} \left( - \frac{\gamma_\phi}{(\Delta k)^2} 
\left| \frac{\partial \widehat{H}}{\partial \widehat{\phi}_t (k_n)} \right|^2 
+ 
\frac{\gamma_\phi}{(\Delta k)^2 \beta}  
\frac{\partial^2 \widehat{H}}{\partial \widehat{\phi}_t (k_n)\partial \widehat{\phi}_t (-k_n)}  \right)
\right] \nonumber \\
=
& - \ud t \int [\ud \Pi] \int [\ud \phi]  \int \ud x\, \rho_t (\{\Pi,\phi\}) 
\left[ 
\gamma_\Pi \left| \frac{\delta H }{\delta \Pi (k)} \right|^2 +
\gamma_\phi \left| \frac{\delta H}{\delta \phi (k)} \right|^2
\right] 
\nonumber \\
& - \ud t \int [\ud \Pi] \int [\ud \phi]  \int \ud x\, \rho_t (\{\Pi,\phi\}) 
\left[ 
\frac{\gamma_\Pi}{\beta} \frac{\delta H}{\delta \Pi (k)} 
\frac{\delta \rho_t ( \{\Pi,\phi\}) }{\delta \Pi (-k)} +
\frac{\gamma_\phi }{\beta} \frac{\delta^2 H}{\delta \phi(k)}  
\frac{\delta \rho_t (\{\Pi,\phi\}) }{\delta  \phi(-k)} 
\right] 
\, ,
\end{align}
where in the last equality we took the limit of continuum, see Appendix \ref{app:continuum},  
and performed some integrations by part. 
Finally, Eq.(\ref{eqn:st2nd}) is the combination of above equations. 


\end{document}